\documentclass[journal=apchd5,manuscript=article, layout=twocolumn]{achemso}

\usepackage{chemformula} 
\usepackage[T1]{fontenc} 

\usepackage{amsmath,amsthm,amssymb}
\usepackage{wasysym} 
\usepackage{textcomp} 

\usepackage{graphicx}
\usepackage[utf8]{inputenc}
\usepackage{lmodern}
\usepackage{microtype}
\usepackage{hyperref}

\usepackage{xcolor}

\definecolor{ao(english)}{rgb}{0.0, 0.5, 0.0}
\definecolor{cadmiumgreen}{rgb}{0.0, 0.42, 0.24}

\author{Elizaveta Gangrskaia}
\affiliation {Photonics Institute, TU Wien, Gusshausstrasse 27-387, A-1040 Vienna, Austria}
\email{elizaveta.gangrskaia@tuwien.ac.at}

\author{Thomas Schachinger}
\affiliation {University Service Centre for Transmission Electron Microscopy (USTEM), TU Wien, A-1040 Wien, Austria}

\author{Christoph Eisenmenger-Sittner}
\affiliation {Institute of Solid State Physics, TU Wien, Wiedner Hauptstraße 8-10, A-1040 Wien, Austria}

\author{Lorenz Gr\"{u}newald}
\affiliation {Institute of Theoretical Chemistry, Faculty of Chemistry, University of Vienna, Währinger Str. 17, A-1090 Vienna, Austria}
\alsoaffiliation{Vienna Doctoral School in Chemistry (DoSChem), Faculty of Chemistry,
University of Vienna, Währinger Str. 42, A-1090 Vienna, Austria}

\author{Sebastian Mai}
\affiliation {Institute of Theoretical Chemistry, Faculty of Chemistry, University of Vienna, Währinger Str. 17, A-1090 Vienna, Austria}

\author{Andrius Baltu\v{s}ka}
\affiliation {Photonics Institute, TU Wien, Gusshausstrasse 27-387, A-1040 Vienna, Austria}
\alsoaffiliation{Center for Physical Sciences \& Technology, Savanoriu Ave. 231, LT-02300 Vilnius, Lithuania}

\author{Audrius Pug\v{z}lys}
\affiliation {Photonics Institute, TU Wien, Gusshausstrasse 27-387, A-1040 Vienna, Austria}
\alsoaffiliation{Center for Physical Sciences \& Technology, Savanoriu Ave. 231, LT-02300 Vilnius, Lithuania}

\author{Alessandra Bellissimo}
\affiliation {Photonics Institute, TU Wien, Gusshausstrasse 27-387, A-1040 Vienna, Austria}
\email{alessandra.bellissimo@tuwien.ac.at}

\title{Probing Optical Magnetic Dipole Transitions in Eu$^{3+}$ using Structured Light and Nanoscale Sample Engineering}


\begin{document}

\begin{abstract}
    At optical frequencies, interactions of the electric field component of light with matter are dominating, whereas magnetic dipole transitions are inherently weak and challenging to access independently from electric dipole transitions.
    However, magnetic dipole transitions are of interest, as they can provide valuable complementary information about the matter under investigation.
    Here, we present an approach which combines structured light irradiation with tailored sample morphology for enhanced and high-contrast optical magnetic field excitation, and we test this technique on ${\text{Eu}^{3+}}$ ions. 
    We generate spectrally tunable, narrowband, polarization-shaped ultrashort laser pulses, which are specifically optimized for the spectral and the spatial selective excitation of magnetic dipole and electric dipole transitions in ${\text{Eu}^{3+}:\text{Y}_2\text{O}_3}$ nanostructures integrated into a metallic antenna. 
    In the presence of the metallic antenna, the excitation with an azimuthally polarized beam is shown to provide at least a 3.0--4.5-fold enhancement of the magnetic dipole transition as compared to a radially polarized beam or a conventional Gaussian beam.
    Thus, our setup provides new opportunities for the spectroscopy of forbidden transitions.
\end{abstract}

\section{Introduction}

Optical spectroscopy studies the fundamental properties of atoms, molecules, and solid-state materials through their interaction with light, e.g., through absorption or emission~\cite{TKACHENKO2006}. 
Light-matter interactions are often described by expanding them into multipolar transitions: electric dipole (ED), magnetic dipole (MD), electric quadrupole, etc~\cite{Craig1984}. 
The ED and MD optical transitions connect different electronic states and occur with probabilities defined by the relevant selection rules~\cite{Cowan1981}. 
At optical frequencies, ED moments in matter interact with the electric field (EF) component of the light approximately $10^5$ times stronger than MD moments interact with the magnetic field (MF) counterpart~\cite{Gawad2012}. 
Most investigations of light-matter interactions in the optical regime have therefore focused on ED transitions driven by the EF, while MD transitions driven by the MF have received comparatively little attention, see, e.g., Refs.~\citenum{Stephens1974, SMAIL2025100544}. 
However, because selection rules forbid certain ED transitions, the role of magnetic interactions in these ``forbidden'' cases becomes particularly important. 
In fact, a direct access to these ``forbidden'' optical transitions could provide valuable insights into the molecular symmetry as well as the electronic and vibronic structure of the investigated atomic / molecular systems, consequently providing additional information on the nature of these photoinduced processes~\cite{Manjavacas2017}.
This highlights the importance of establishing a ``MD-based spectroscopy'', complementary to traditional ED-based optical spectroscopy.

Typically, probing MD transitions is challenging, since they are intrinsically weak and can be easily obscured by spectrally adjacent ED transitions. 
The magnetic response of a spectroscopic target can be strengthened by employing plasmonic\cite{Hrton2020, Abdellaoui2015, Hussain2015, Darvishzadeh-Varcheie2017, Yang2022, Choi2016} or dielectric antennas~\cite{Calandrini2018, Baranov2017, Wiecha2019, Bonod2019, Sugimoto2021, Vaskin2019, Albella2013, Cheng2021, Liang2019}, leading to the local enhancement of the optical MF.
However, such micro- and nanostructures often lead to comparable or even stronger EF enhancement in their vicinity~\cite{Hrton2020}. 
Therefore, the detection of high-contrast MF-MD interactions requires both the enhancement of the MF and the suppression of the EF component in a spatial region of interest.

One of the approaches allowing spatial separation of the ED/MD resonances is placing a thin sample into a standing wave, which, depending on the sample position leads to the selectivity of the EF/MF excitation~\cite{Fang2016,Reynier2023}. 
An alternative concept to enhancing the MF involves shaping the excitation light in terms of both spatial distribution and polarization.
Among examples of structured light there are cylindrical vector beams with azimuthal or radial polarization and doughnut-shaped spatial profiles~\cite{Zhan2009}. 
Radially polarized beams (RPBs) feature a spatially isolated axial EF component.
In azimuthally polarized beams (APBs), the MF has a spatially isolated longitudinal component along the propagation direction, with the transverse EF and MF components vanishing near the beam center.
A tightly focused APB forms a region of MF dominance that is suitable for selective excitation of MD transitions~\cite{Kasperczyk2015,Banzer2021,Veysi2016}.

So far, only a few, mostly theoretical works have been devoted to structured beam excitation combined with magnetic optical antennas~\cite{Blanco2019,Darvishzadeh-Varcheie2019,Martín-Hernández2024, Manna2017,Guclu2016,Das2017}. 
Although previous studies have explored the interaction between structured light and magnetic optical resonances---particularly using APBs to excite magnetic dipole modes in dielectric Mie resonators, and to detect magnetic near-fields via photoinduced force microscopy\cite{Zeng_MieResonator_2018}---the experimental combination of ultrafast APBs at visible frequencies with metallic nanoantennas remains largely unexplored. 
Recent simulations suggest that tailored metallic nanoantennas can significantly enhance the longitudinal magnetic field component of APBs, achieving field strengths of up to several tens of Tesla\cite{Blanco2019, Martín-Hernández2024}.

\begin{figure}[tb]
    \centering
    \includegraphics[width=1.0\linewidth]{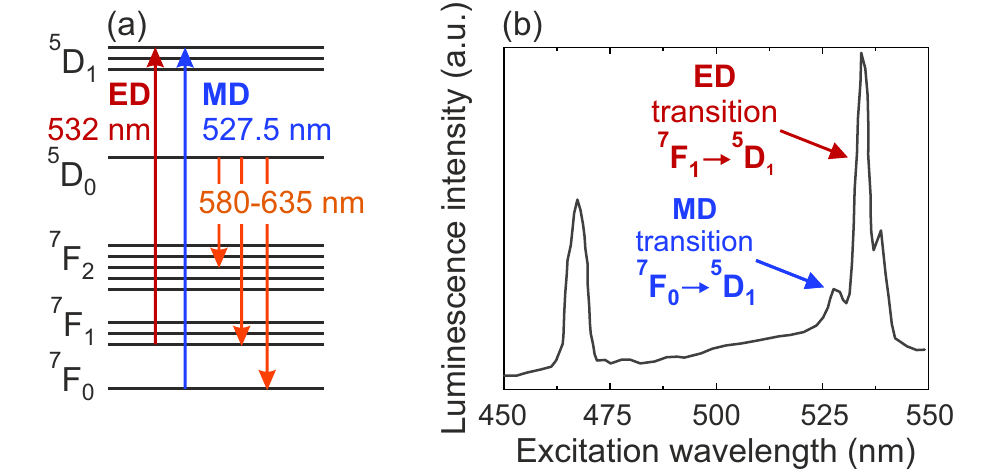}
    \caption{(a) Simplified energy level structure of ${\text{Eu}^{3+}}$ ions in the ${\text{Y}_2\text{O}_3}$ host matrix. (b) Photoluminescence excitation spectrum of 5~mol\% ${\text{Eu}^{3+}:\text{Y}_2\text{O}_3}$ nanoparticles for a collection at wavelength $\lambda=$ 612 nm. The data is adapted from Ref \cite{PACKIYARAJ2014}.}
    \label{fig:1intro}
   \end{figure}

\begin{figure*}[t]
    \centering
    \includegraphics[width=1\linewidth]{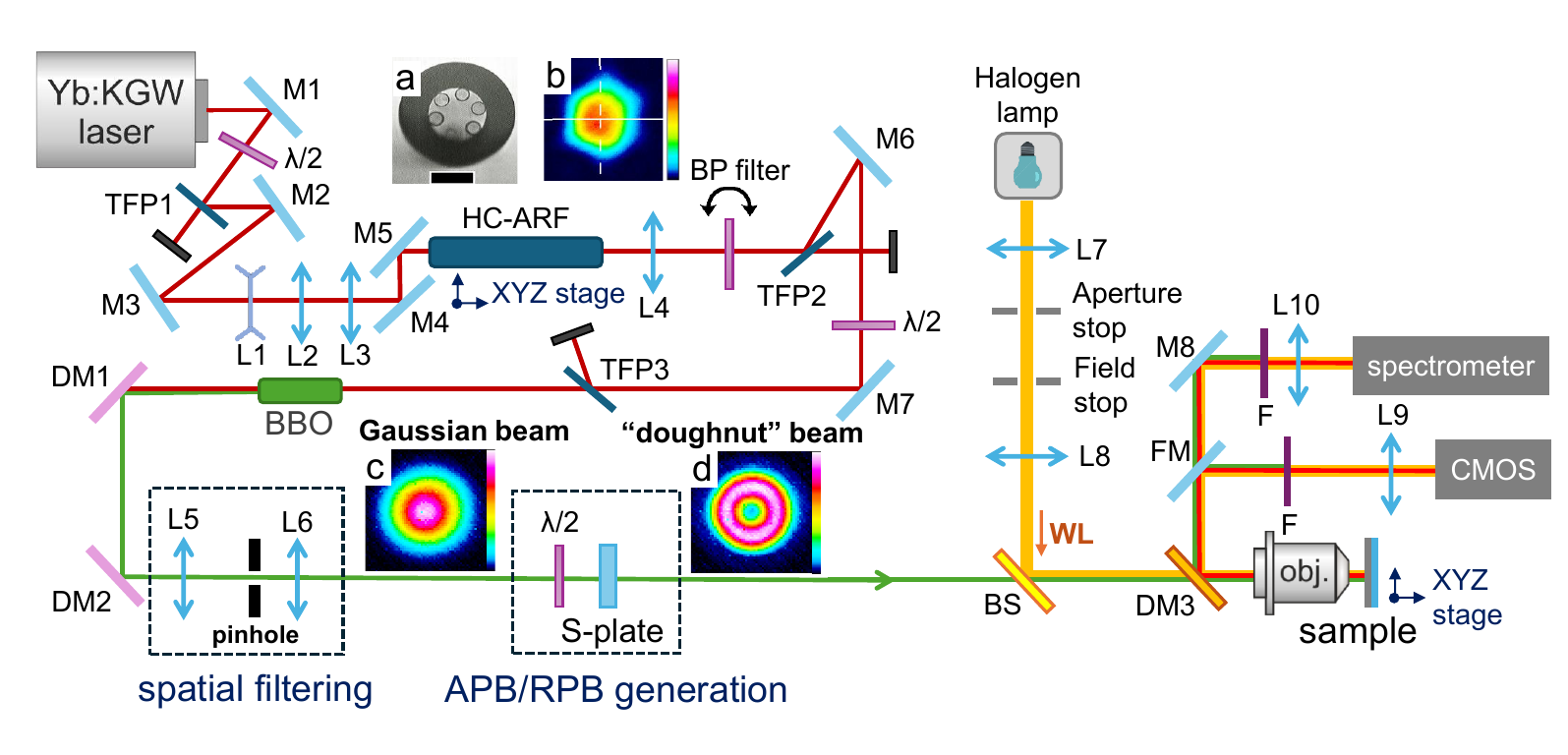}
    \caption{Schematic of the experimental setup. M1-M8: mirrors; $\lambda$/2: half-waveplates; TFP: thin-film polarizers; L1-L10: lenses; HC-ARF: hollow-core anti-resonant fiber; BP: bandpass filter; BBO: $\beta$-barium borate nonlinear crystal; DM1-3: dichroic mirrors; BS: pellicle beam splitter; WL: white light; FM: flipped mirror; obj.: 40$\times$ Olympus plan achromat objective; CMOS: color 1.6MP camera; F: longpass/bandpass filters. Inset (a): microscope image of the HC-ARF input facet. The size of the scale bar is 50~$\mu$m. Inset (b): laser beam profile at the fiber output. S-plate is a polarization converter transforming input Gaussian beam (inset (c)) into a ``doughnut'' beam (inset (d)). }
    \label{fig:2schematic}
\end{figure*}

Building upon these insights, we present an experimental platform that integrates a fluorescent nanostructure with a $\mu$m-sized metallic antenna. 
This configuration, when illuminated with spectrally tunable APBs, leads to a substantial enhancement of the magnetic dipole excitation, offering a test-bed for emerging magnetic dipole-based spectroscopy.

We chose ${\text{Eu}^{3+}}$ ions as the spectroscopic target material, as they are well-studied and exhibit inherently strong intra-4$f$ shell MD transitions in the visible spectral range~\cite{Dodson2012}. 
In lanthanide ions, such as Eu or Er, the partly filled 4$f$ electron shell is shielded by filled ${5\text{s}^2}$ and ${5\text{p}^6}$ orbitals, resulting in sharp intra-4$f$ transitions (see Fig.~\ref{fig:1intro}) with high quantum efficiency, favoring the observation of MD transitions~\cite{GORLLER1998,Edelstein1979}.
While intra-4$f$ ED transitions in isolated lanthanide ions are strictly forbidden by the parity selection rule, this restriction is partially lifted in a crystalline medium, where embedded ${\text{Eu}^{3+}}$ ions exhibit ``induced'' ED transitions that, though weaker than typical allowed ED transitions, are comparable in strength to MD transitions~\cite{Binnemans2015}.
Specifically, the ${\text{Eu}^{3+}:\text{Y}_2\text{O}_3}$ compound features narrow absorption bands centered at 527.5~nm (MD transition ${^7\text{F}_0} \rightarrow {^5\text{D}_1}$) and at 532~nm (ED transition ${^7\text{F}_1} \rightarrow {^5\text{D}_1}$). 
Since the MD transition is driven by the optical MF component and the ED transition, by contrast, is driven by the EF component of light, the spectral separation of $\sim$5~nm between the two transitions provides an additional degree of selectivity, allowing proper testing of the MF/EF spatial isolation achievable in our excitation scheme. 
However, we emphasize that because the enhanced MF is spatially isolated from the EF, our MD-exclusive approach is capable of addressing weak MD transitions that spectrally overlap with ED transitions, as is generally the case in most systems. 

\section{Results and discussion}

\subsection{Generation of narrowband wavelength-tunable APBs}

Spectrally selective excitation of either MD or ED transition in ${\text{Eu}^{3+}}$ ions requires a tunable laser source that operates at least in the range of 527--532~nm and is narrowband enough to avoid crosstalk between the magnetic and electric excitation channels. 
In this section, we present an optical setup specifically tailored for this purpose. 
We exploit ultrashort pulses to achieve spectral tunability using nonlinear optical processes.
To tune the wavelength, we employ stimulated rotational Raman scattering (SRRS) and second-harmonic generation (SHG) in a long crystal, which leads to ``spectral focusing'' and, consequently, spectral narrowing. 
APBs are produced by using an S-waveplate.

\subsubsection{Tunable spectral red shift in a hollow-core fiber}

The front end of the optical setup presented in Fig.~\ref{fig:2schematic} is a 20\,kHz repetition rate Yb:KGW laser system (Pharos, Light Conversion) delivering 200~$\mu$J, 350~fs pulses centered at 1030~nm. 
For the generation of a narrowband, tunable radiation in the 527–532~nm spectral window required for selective excitation of the MD and ED transitions in the ${\text{Eu}^{3+}:\text{Y}_2\text{O}_3}$ compound, we exploit SRRS as the first spectral-shaping step. 
To this end, the collimated laser beam was enlarged with a Galilean beam expander (lenses L1 and L2) and focused with the lens L3 onto the end face of a free-standing 75\,cm long, hollow-core antiresonant fiber (HC-ARF). 
The inner core diameter of the HC-ARF is 30~$\mu$m, the cladding consists of 6 hollow glass cylinders with $\sim$19~$\mu$m diameter and the wall thickness of $\sim$330 nm (Fig.~\ref{fig:2schematic}a). 
While propagating through the fiber, the ultrashort laser pulses interact with air molecules (mainly $\text{N}_2$ and $\text{O}_2$) confined within the core and when the pump peak intensity exceeds a threshold, the SRRS process starts. 
During SRRS, the pump photons transfer energy to molecular rotations, which results in the generation of lower energy Stokes photons. 
Further propagation of the generated photons leads to a cascaded generation of next-order Stokes components. 
As a result, the pulses transmitted through the HC-ARF filled with air obtain a broadened and red-shifted spectrum. 

\begin{figure}[tb]
    \centering
    \includegraphics[width=0.75\linewidth]{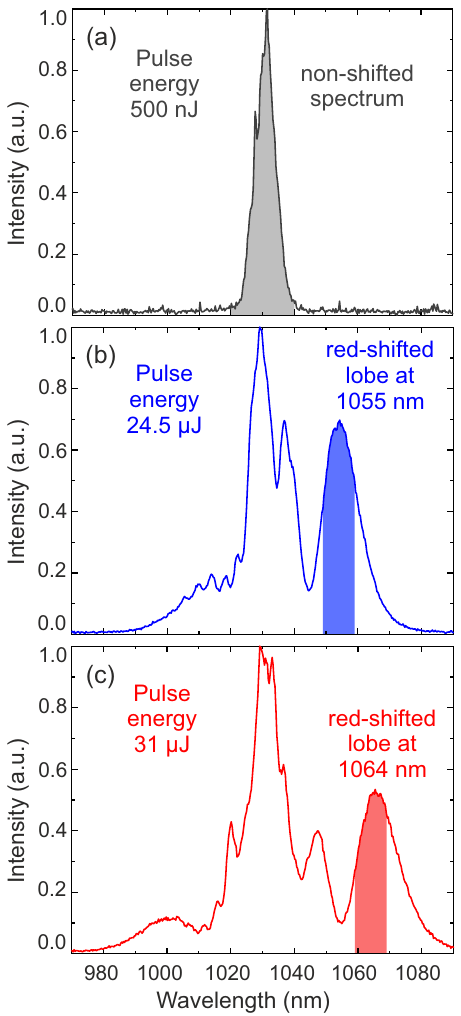}
    \caption{Spectral output obtained through the SRRS process after the hollow-core fiber.
    (a) At 500\,nJ input pulse energy, no spectral shift is observed.
    (b) A 24.5~$\mu$J input produces a red-shifted lobe centered at 1055~nm.
    (c) A 31~$\mu$J input produces a red-shifted lobe centered at 1064~nm, instead.
    Blue and red shaded regions indicate the spectral range selected with the bandpass filter.}
    \label{fig:3shift}
   \end{figure}

The self-phase modulation and SRRS-induced spectral broadening and shift ${\Delta\omega}$ depend on the fiber dimensions (length $L$ and section area $A$), pulse peak power $P$, laser central wavelength $\lambda_0$, and gas type and pressure as follows~\cite{Carpeggiani2020}:
\begin{equation}
   \Delta\omega \propto \dfrac{L}{A} \cdot P \cdot \dfrac{\kappa_2 \cdot p}{\lambda^2_0},
   \label{eq:DE_uncoupled}
\end{equation}
where $\kappa_2$ is the ratio between the nonlinear index coefficient and the gas pressure $p$. 
The spectral shift at the fiber output was controlled by only adjusting the pulse energy, which directly affects the pulse peak power $P$. 
The pulse energy was controlled by a variable attenuator consisting of a half-wave plate and a thin-film polarizer. 
The spectra obtained at different pulse energy are presented in Fig.~\ref{fig:3shift}. 
For low pulse energy (500~nJ), no spectral shift is observed (Fig.~\ref{fig:3shift}a). 
For the input fs-pulses of 24.5~$\mu$J energy, the red-shifted lobe in the output spectrum is centered at 1055~nm (Fig.~\ref{fig:3shift}c). 
The red-shifted lobe was spectrally separated from the pump light by tilting an interference 1100-nm bandpass (BP) filter with a bandwidth of 10~nm FWHM (blue tinted area in Fig.~\ref{fig:3shift}b). 
To tune the rightmost spectral lobe to 1064~nm, the input pulse energy was increased to 31~$\mu$J, and the BP filter was tilted accordingly (Fig.~\ref{fig:3shift}c).

\begin{figure}[tb]
    \centering
    \includegraphics[width=0.75
    \linewidth]{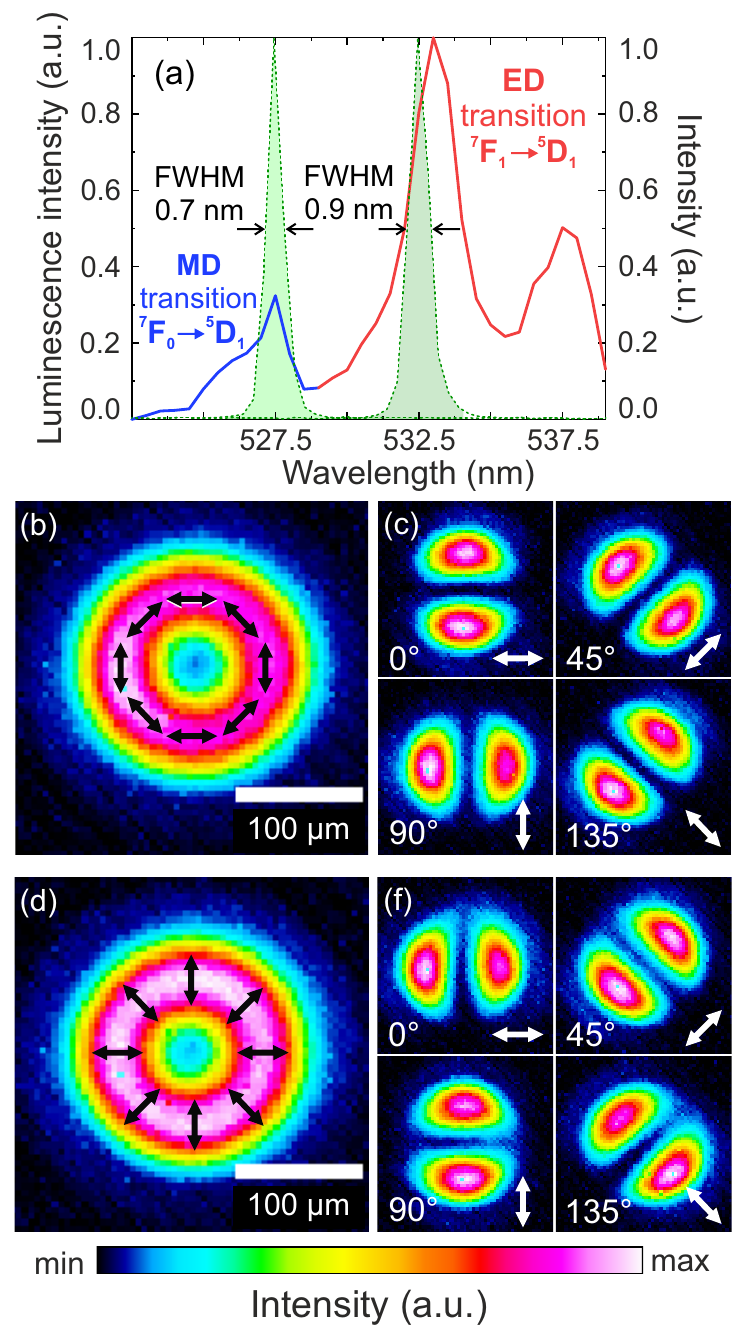}
    \caption{
    (a) Excitation spectrum of ${\text{Eu}^{3+}:\text{Y}_2\text{O}_3}$ thin film measured by collecting the integrated emission signal in the 580--620~nm spectral range. Distinct peaks (see also the energy level diagram in Fig.~\ref{fig:1intro}) correspond to the MD transition (solid blue line) and the ED transition (solid red line). Narrowband spectra of SH pulses tuned to 527.5~nm (532.5~nm) are shown as green filled areas.
    CCD image of the (b) APB and (d) RPB intensity profiles, with black arrows indicating the polarization direction. 
    Laser beam profiles observed after passing (c) APB and (f) RPB through a Glan-Taylor polarizer placed in front of the CCD with the transmittance axis oriented at $0^{\circ}$, $45^{\circ}$, $90^{\circ}$, and $135^{\circ}$ (highlighted by white arrows).}
    \label{fig:4green}
\end{figure}

\subsubsection{SHG and spectral narrowing in a long nonlinear crystal}

In the next stage, the red-shifted pulses were focused into a 2.5~cm-long $\beta$-barium borate (BBO) crystal for the SHG. 
Frequency-doubling of 1055~nm pulses resulted in 527.5~nm wavelength, corresponding to the MD transition in ${\text{Eu}^{3+}}$ ions (Fig.~\ref{fig:4green}a). 
To reach the ED resonance at 532~nm, the fundamental wavelength was shifted to 1064~nm and the phase-matching angle of the BBO crystal was adjusted accordingly. 
During the SHG process in the long BBO crystal, the broadband fundamental frequency (FF) pulses were ``spectrally compressed'' into narrowband (FWHM$<$1~nm) SH pulses. 
The principle of the ``spectral compression'' is based on the high group velocity mismatch between the FF and SH pulses interacting in a long nonlinear crystal~\cite{Marangoni2007}.

The normalized spectra of SH pulses overlapping with the excitation spectrum of the ${\text{Eu}^{3+}}$ ions are presented in Fig.~\ref{fig:4green}a. 
The excitation spectrum was measured by scanning the excitation wavelength with a step of 0.5~nm and collecting the integrated emission signal in the 580--620~nm spectral range from the whole illuminated area of a thin 2.6\% ${\text{Eu}^{3+}}$-doped ${\text{Y}_2\text{O}_3}$ film of $\sim$400~nm thickness, deposited using radio frequency magnetron sputtering.

\subsubsection{Generation and characterization of APBs and RPBs}

The obtained spectrally tunable narrowband pulses with a Gaussian spatial intensity profile were converted into APBs or RPBs by employing a commercially available S-waveplate (Altechna)~\cite{Beresna2011}. 
Rotation of the half-wave plate placed before the S-waveplate by $\pm$45$^\circ$ leads to the generation of either azimuthally or radially polarized beams. 
We analyzed the polarization state of the generated beams by inserting a Glan--Taylor polarizer and registering the transmitted beam profile with a CCD camera at different orientations of the polarizer. 
After passing through the polarizer, the APB (Fig.~\ref{fig:4green}b) or the RPB (Fig.~\ref{fig:4green}d) transforms into a distinctive two-lobe pattern that rotates consistently with the rotation of the polarizer (Fig.~\ref{fig:4green}c,f).
Due to spectral and polarization tunability, the generated laser pulses permit selective excitation of the MD or ED transition in the ${\text{Eu}^{3+}:\text{Y}_2\text{O}_3}$ spectroscopic target. 

\subsection{Enhancement of the local MF by metallic antennas}

In addition to the spectral and spatial selectivity, the enhancement of the optical MF is highly desirable. 
For this purpose, we use a conductive circular aperture as a ``magnetic'' antenna that locally enhances the longitudinal MF component of an incident APB.
In the following text, the terms ``antenna'' and ``aperture'' are used interchangeably.
As initially proposed by Blanco et al.~\cite{Blanco2019}, focusing an APB onto a ring-shaped metallic aperture results in an MF enhancement by a factor of $\sim 6$ as compared to the unapertured case. 
An APB features an EF that oscillates around the beam perimeter (Fig.\ref{fig:4green}c) driving electronic ring currents at the edge of a metallic aperture and producing an ultrafast, ``tip''-shaped MF at the centre of this aperture.
This magnetic tip oscillates at the laser frequency along the beam propagation axis and extends over a length of about 1~$\mu$m.
Mart\'{\i}n-Hern\'andez et al.\cite{Martín-Hernández2024} later reported an expanded set of numerical simulations that showed how the antenna's shape, thickness, and geometrical parameters affect the strength, size, and position of the MF ``tip''. 
In the case of the simplest antenna shape (a flat cylindrical aperture), the radius of the aperture must coincide with the radius of maximum intensity of the APB $(\rho_0)$ to maximize the electronic currents induced at the edge of the aperture and thus optimize the MF enhancement~\cite{Blanco2019}. 

Besides the MF intensity, another critical parameter for MD-exclusive spectroscopy is a high local contrast between the MF and EF, which reflects the spatial separation of the MF component.
The simulations predict~\cite{Blanco2019} that the presence of the aperture selectively enhances the longitudinal MF component without affecting the transverse EF, which results in an increased MF/EF contrast for the apertured configuration as compared to the unapertured case.
For the studied antennas, the lateral extent of the region with high MF/EF intensity contrast does not significantly depend on the geometrical configuration of the antenna and is mainly linked to the laser wavelength. 
For the excitation wavelength in the range of 527--532~nm, the transverse region supporting at least a 10-fold intensity contrast (defined as $c^2|B|^2/|E|^2$) is limited to about 100~nm~\cite{Martín-Hernández2024}. 

\subsection{Fabrication of a ${\text{Eu}^{3+}:\text{Y}_2\text{O}_3}$ nanostructure integrated into a magnetic antenna}

The spectroscopic sample must be placed in the region of the MF dominance, but its lateral size must remain sufficiently small to avoid unwanted excitation by the EF. 
Therefore, the ${\text{Eu}^{3+}:\text{Y}_2\text{O}_3}$ sample should be a nanostructure of $\sim$100~nm diameter, precisely positioned in the center of the antenna. 
Since the precise placement of nanoparticles relative to pre-fabricated antennas is highly challenging, we developed a fabrication procedure that enables the direct integration of the nanostructures into the antennas. 
The procedure combines a ${\text{Eu}^{3+}:\text{Y}_2\text{O}_3}$ thin film deposition by radio frequency magnetron sputtering, metal film depositions by direct-current (DC) magnetron sputtering, and milling of the nanostructures by focused ion beam (FIB) etching.

\begin{figure}[tb]
    \centering
    \includegraphics[width=1.0\linewidth]{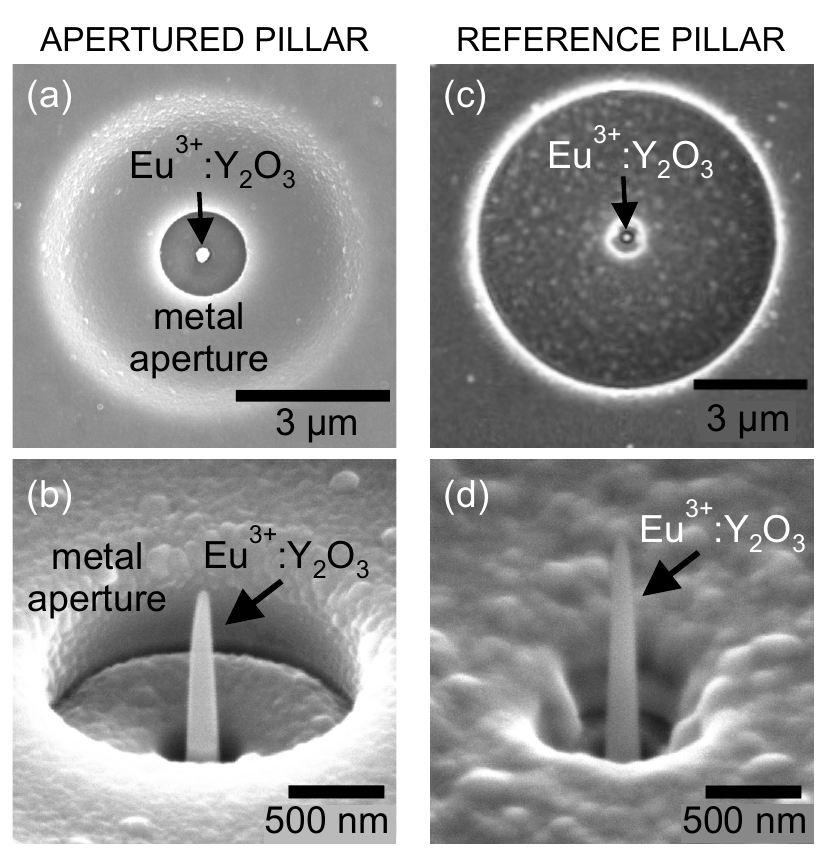}
    \caption{Scanning electron microscope (SEM) images of (a,c) the top view and (b,d) the $52^{\circ}$ tilt view of the apertured and unapertured (reference) ${\text{Eu}^{3+}}$ nanostructures, correspondingly. 
    }
    \label{fig:5samples}
\end{figure}

The 1.2~$\mu$m-thick  ${\text{Eu}^{3+}:\text{Y}_2\text{O}_3}$ film deposited onto an ITO (indium tin oxide)-coated glass substrate features a smooth, uniform structure suitable for the high precision nanopatterning process. 
Using a FIB machine, we milled the ${\text{Eu}^{3+}:\text{Y}_2\text{O}_3}$ film to form a $\diameter1.8~\mu$m aperture with a central pillar, then deposited a 220~nm Al film, and selectively removed it from the pillar, yielding a $\diameter1.6~\mu$m metallic antenna surrounding the isolated ${\text{Eu}^{3+}:\text{Y}_2\text{O}_3}$ pillar (the ``apertured nanopillar'' in Fig.~\ref{fig:5samples}a,b).
The pillar is a 1~$\mu$m-tall truncated cone with $\sim$80~nm diameter at the top and $\sim$160~nm at the pedestal. 
The elongated shape of the pillar is beneficial for maximizing the luminescence signal, since the enhanced MF ``tip'' forms in the aperture's center, along the optical axis (perpendicular to the sample surface). 

To identify the contribution of the metal antenna, we fabricated a unapertured pillar of similar size (``reference pillar'' in Fig.~\ref{fig:5samples}c,d).
The reference pillar is surrounded by an $\diameter$8~$\mu$m opening, which is too large to be affected by the APB excitation beam tuned to the size of  the  $\diameter$1.6~$\mu$m aperture.
A metal layer thinned down to 125~nm was retained around the isolated pillar to cover the surrounding luminescent film.
A small clearance around the pillar ($\sim$400~nm in size) created during the pillar shaping process has an irregular, non-circular shape with rough edges and is not considered a functional antenna. 
Therefore, no observable MF enhancement is expected in the reference configuration. 
Full details on the sample preparation are given in the Methods section.

\subsection{Numerical simulations}

\begin{figure}[tb]
    \centering
    \includegraphics[width=1.0\linewidth]{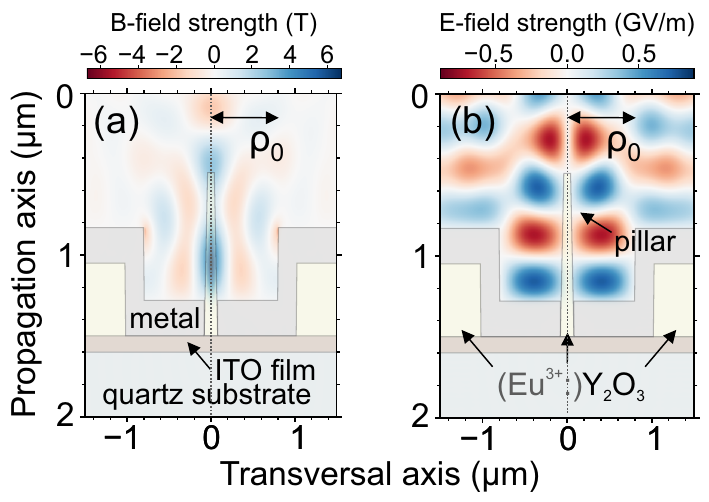}
    \caption{Numerically simulated distribution of the (a) longitudinal MF and (b) azimuthal EF components in the vicinity of the apertured nanopillar for the APB excitation. 
    In the drawing, the central nanopillar (yellow) is surrounded by the metal layer (grey). 
    The metal layer is deposited on bulk ${\text{Eu}^{3+}:\text{Y}_2\text{O}_3}$ (yellow) and the sample is attached to a quartz substrate (green) with an intermediate ITO film (brown).
    In the simulation, rotational symmetry around the vertical dotted line is obeyed.
    }
    \label{fig:6theory}
\end{figure}

To verify and illustrate the enhancement and isolation of the MF for our setup, we performed numerical finite-difference time-domain (FDTD)\cite{2013_taflove_Advances,2005_taflove_Computational} simulations with the open-source code MEEP\cite{2010_oskooi_Meep}.
In contrast to the more approximate particle-in-cell (PIC)\cite{2002_fonseca_OSIRIS} framework within our previous study\cite {Martín-Hernández2024}, the FDTD simulations allow incorporating intrinsic material parameters---specifically, the frequency- and intensity-dependent electric permittivity and magnetic permeability---which allow for a more realistic description of the metallic and insulating materials.
Our results confirm that a strong MF is build up at the pillar surrounded by the metallic aperture, reaching up to 6.7~T (Fig.~\ref{fig:6theory}a). 
For comparison, a Gaussian beam with the same beam parameters, i.e. featuring the same minimal beam waist and intensity, yields a maximum MF of roughly 0.5~T, which also lies on the beam propagation axis.
In addition to requiring strong MF intensities, the lack of any parasitic EF contribution---as is demonstrated here---is equally necessary to obtain a clean spectroscopic signal of weak MD transitions.
We can observe the distinct build-up of a magnetic needle at the propagation axis (Fig.~\ref{fig:6theory}a) while the electric field is absent in the same area (Fig.~\ref{fig:6theory}b).
To assess the quality of the MF isolation at the nanopillar, we computed the integrated MF and EF dynamical energy densities in time and over the spatial volume of the ${\text{Eu}^{3+}:\text{Y}_2\text{O}_3}$ nanopillar.\cite{1999_jackson_Classical, 2014_welters_Speedoflight}
The estimated ratio of the integrated MF and EF dynamical energy densities, i.e., the relevant intensity contrast, is approximately 6.4 times higher for APB excitation than that for Gaussian beam excitation.

\subsection{Experimental evaluation of the magnetic antenna for selective MD excitation}

The prepared ${\text{Eu}^{3+}:\text{Y}_2\text{O}_3}$ nanopillars were excited with narrowband, spectrally tunable APBs, RPBs, and Gaussian beams to test their respective performance.
By measuring luminescence excitation spectra, we estimated the MF/EF intensity contrast in the region where the luminescent nanostructure is located. 
To identify the MF enhancement, provided by the metallic antenna, we compared the performance with that of the reference unapertured case.

\begin{figure}[tb]
    \centering
   \includegraphics[width=1.0\linewidth]{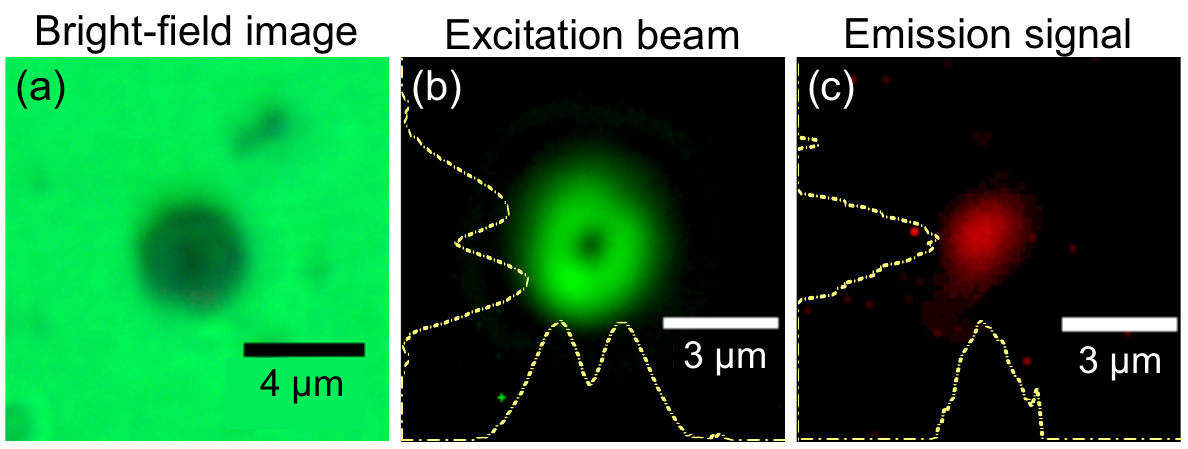}
    \caption{(a) Bright-field image of the sample.  (b) Intensity profile of a focused APB. (c) Luminescence intensity pattern collected from the ${\text{Eu}^{3+}}$ nanopillar excited with APB. Yellow dashed lines represent integrated vertical and horizontal intensity profiles. 
    }
    \label{fig:7micro}
\end{figure}

We constructed an optical microscope based on a single 40$\times$ Olympus plan achromat microscope objective (Thorlabs RMS40X) to visualize and locally excite the fabricated nanostructures. 
For the visualization, the sample was uniformly illuminated with a tungsten halogen lamp (OceanOptics LS-1) through a 45/55 pellicle beam splitter (Fig.~\ref{fig:2schematic}). 
The bright field image of the sample is shown in Fig.~\ref{fig:7micro}a as the profile of the focused APB is presented in Fig.~\ref{fig:7micro}b. 
The size of the APB and RPB (beam waist $w_0=1.2$~$\mu$m, $\rho_0=\dfrac{w_0}{\sqrt{2}} \thickapprox 0.8$ $\mu$m) was adjusted to the size of the aperture. 
The Gaussian beam was obtained by removing the S-waveplate from the beam path. 
The luminescence signal emitted by the $\text{Eu}^{3+}$ ions (Fig.~\ref{fig:7micro}c) was collected using the same microscope objective. 
A dichroic mirror (DM3 in Fig.~\ref{fig:2schematic}) transmitted the green excitation pulses and reflected the collected luminescence. 
The residual green light was either attenuated for monitoring the beam profile, or completely blocked by a BP filter ($\lambda_c$=600~nm, FWHM=40~nm) for the luminescence measurements. 
The bright field image of the sample, the profile of the excitation beam, and luminescence patterns were observed by a Zelux 1.6 MP color CMOS camera. 

The extraction of information about the achieved MF/EF contrast is possible since ${\text{Eu}^{3+}}$ ions feature spectrally separated MD and ED transitions.
Specifically, as it is illustrated in Fig.~\ref{fig:1intro}, the excited ${^5\text{D}_1}$ state can be populated through two different excitation pathways: the MD transition at 527.5~nm from the ground level ${^7\text{F}_0}$ (by the optical MF component) or the ED transition at 532~nm from the thermally populated state ${^7\text{F}_1}$ (by the optical EF component). 
Once the ${^5\text{D}_1}$ is populated, the ${^5\text{D}_0}$ state is populated by internal conversion, and the subsequent spontaneous emission from that state is governed solely by intrinsic radiative properties and is independent of the excitation mechanism~\cite{Kasperczyk2015}.
Consequently, any variation in the emission intensity results from the different excitation rates of MD and ED transitions. 
In the linear, low-power regime the excitation rate for the MD (ED) transition is proportional to the intensity of the exciting MF (EF) component. 
Changes in the relative intensities of the MD and ED peaks in the excitation spectrum reflect the corresponding variation in the MF/EF intensity contrast in the given excitation scheme. 

\begin{figure}[tb]
    \centering
    \includegraphics[width=0.9\linewidth]{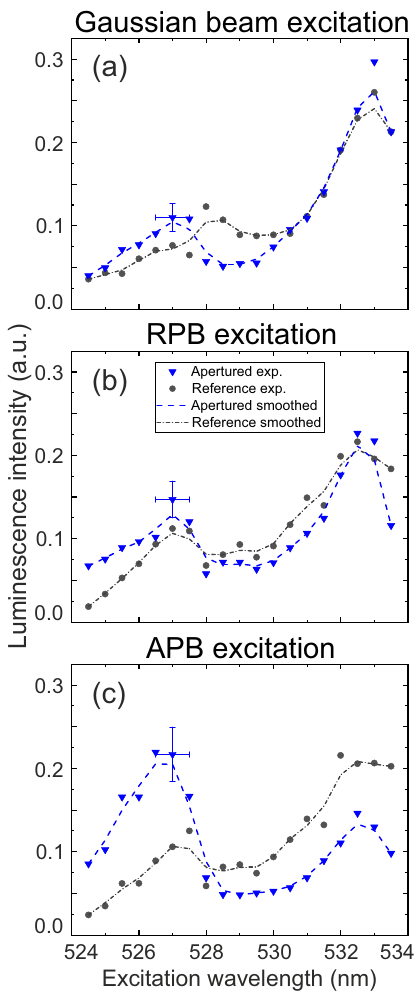}
    \caption{Excitation spectra of the apertured ${\text{Eu}^{3+}:\text{Y}_2\text{O}_3}$ nanopillar (blue triangles) and the reference unapertured ${\text{Eu}^{3+}:\text{Y}_2\text{O}_3}$ nanopillar (black circles) excited with (a) Gaussian beam, (b) RPB, and (c) APB. All curves are normalized to the total area and smoothed. The symbols represent the raw data points, while the dashed lines of the same color show the smoothed data. For clarity, representative error bars are shown at the data point 527~nm.}
    \label{fig:8spectra}
\end{figure}

The obtained luminescence excitation spectra are presented in Fig.~\ref{fig:8spectra}. 
We scanned the excitation wavelength of the Gaussian beam (Fig.~\ref{fig:8spectra}a), RPB (Fig.~\ref{fig:8spectra}b), and APB (Fig.~\ref{fig:8spectra}c) in the range 524.5--533.5~nm in 0.5~nm steps and collected the integrated luminescence in the spectral range $\lambda=$ 580--620~nm. 
For comparison of the shapes of the measured luminescence excitation spectra, the measured curves were normalized to the total area under the curves and subsequently smoothed using binomial smoothing.
We evaluated the MF/EF intensity contrast by calculating the ratio of the MD to ED transition peaks in the excitation spectra. 
It is worth mentioning that the accuracy of the excitation wavelength measurements was affected by the spectral resolution of the spectrometer (OceanOptics USB2000+), determined by the distance between the adjacent pixels corresponding to a 0.36~nm spectral interval (see the estimated error bars for wavelengths in Fig.~\ref{fig:8spectra}).
The maxima in the spectra near 527.5~nm and 532~nm were taken as corresponding to the MD transition and the ED transitions, respectively.

In the case of excitation by the Gaussian beam (Fig.~\ref{fig:8spectra}a), the ratio between the emitted signal for the MD and ED transitions does not exceed 0.4 for either the apertured (blue triangles) or the reference (black circles) cases. 
A similar ratio is observed in the excitation spectrum of another thin film prepared using the same method (Fig.~\ref{fig:4green}b).
Furthermore, we illuminated the nanopillars with the RPB (Fig.~\ref{fig:8spectra}b), which has a doughnut-shaped intensity profile similar to that of the APB, but exhibits a different polarization pattern (Fig.~\ref{fig:4green}b,d) and features a longitudinal EF component on the optical axis, in contrast to the MF component in APB. 
No significant improvement in the MF/EF contrast is observed for the apertured nanopillar compared to the reference, indicating that, as expected, the antenna is not activated by RPB excitation.
Under APB excitation (Fig.~\ref{fig:8spectra}c), the presence of the aperture leads to the threefold increase in ratio between the MD and ED peaks compared to the unapertured case, indicating a selective enhancement of the MF component.
Moreover, combining APB excitation with the metallic aperture boosts the MF/EF contrast by a factor of 4.5 relative to Gaussian beam excitation.
The experimentally estimated enhancement factor of 4.5 is in reasonable agreement with the simulated value of 6.4.

As shown in Ref.~\citenum{Kasperczyk2015}, an APB by itself is well-suited for the selective excitation of the MD transition in ${\text{Eu}^{3+}:\text{Y}_2\text{O}_3}$ nanoparticles when tightly focused ($w_0 \thickapprox$230 nm). 
The ratio between the maximum longitudinal MF and maximum transverse EF scales inversely with the beam waist $w_0$: $\dfrac{\text{MF}(w=0)}{\text{EF}(w=\rho_0)}=\dfrac{0.74\lambda}{w_0}$~\cite{Veysi2016}. 
In this work, we employed a $\sim$5-times larger APB resulting  in a reduced longitudinal MF and MF/EF intensity ratio compared to the tightly focused case.
Given the estimated sample displacement from the vibrations of the optical table (up to $\sim$350~nm), along with the finite size of the nanostructure and minor alignment variations, the transverse EF and MF components, due to their partial spatial overlap with the nanopillar, play a more prominent role than the non-enhanced longitudinal MF component. 
This helps explain the smaller distinction observed between the APB and RPB excitation spectra for the reference pillar, as well as the slightly lower experimental MF/EF contrast achieved with the aperture.

\section{Conclusion}

In conclusion, we have experimentally demonstrated a working prototype of a magnetic optical antenna capable of selectively enhancing MD transitions. 
By exciting a ${\text{Eu}^{3+}:\text{Y}_2\text{O}_3}$ nanopillar, surrounded by a cylindrical metallic aperture, with spectrally tunable, narrowband APB ultrashort laser pulses, we achieved a 4.5-fold enhancement in relative luminescence intensity for the ${^7\text{F}_0} \rightarrow {^5\text{D}_1}$ MD transition compared to excitation with a conventional Gaussian beam. 
This enhancement of the MF/EF contrast is in fair agreement with numerical simulations and confirms the effectiveness of our design in boosting the optical MF while maintaining spatial isolation from the EF component. 
These results establish a viable pathway for the controlled and efficient excitation of weak MD transitions in solid-state systems, even under realistic experimental conditions involving moderate focusing and mechanical instabilities.

In future work, we will extend this study by exploring more sophisticated antenna geometries to systematically investigate their impact on the spatial distribution and relative enhancement of the MF and EF components.
Particular attention will be given to how aperture diameter, shape, and the choice of metal affect the MF/EF intensity contrast. 
Once an optimal magnetic antenna design is identified and fully characterized using the ${\text{Eu}^{3+}:\text{Y}_2\text{O}_3}$ testbed system, the demonstrated MD-exclusive spectroscopic approach can be readily applied to other systems exhibiting weak MD transitions that spectrally overlap with ED transitions. 
The demonstrated laser setup is easily adaptable to other spectroscopic targets by tuning the SRRS conditions, such as gas type and pressure inside the HC-ARF~\cite{Carpeggiani2020}. 
Furthermore, the presented magnetic antennas operate effectively over a broad wavelength range. 
In summary, the combination of tailored nanofabrication with structured light excitation in our setup enables the selective enhancement of magnetic dipole transitions, offering a powerful and adaptable toolkit for future spectroscopic investigations across a broad range of optical systems.

\section{Methods}

\subsection{Thin film deposition and nanopatterning} \label{sec:fabrication}

We deposited a 1.2~$\mu$m-thick ${\text{Eu}^{3+}:\text{Y}_2\text{O}_3}$ film onto a 20x15~${\text{mm}^2}$ ITO-coated soda-lime glass substrate (Ossila) by radio frequency magnetron sputtering using a 2 inch diameter target lightly pressed from 4\% ${\text{Eu}^{3+}:\text{Y}_2\text{O}_3}$ micropowder (Sigma--Aldrich).
Sputtering was conducted at 350◦C in an Ar atmosphere (3~Pa), at 80~W radio frequency power with the DC bias of the discharge ranging from 120 to 160 V. The distance from target to substrate was 5~cm, and the deposition time was 480~min at a deposition rate of 2.5~nm/min. 
The fabricated films were plasma-cleaned for 30--60 s using He plasma (Gala Instruments PlasmaPrep5) and sputter coated with a 5~nm Au:Pd layer (Quorum QT150TS) in order to enhance the samples' conductivity using a target to substrate distance of 4~cm, a sputter current of 150~mA, a sputter time of 5~s and an Ar pressure of 0.5~Pa.
The 220 nm~Al thin-film acting as a metal antenna was DC-magnetron sputtered using the QT150TS at 0.5~Pa Ar pressure, 150~mA sputtering current, 380~s sputtering time and a base pressure of $7\times10^{-6}\,\text{mbar}$ with a target to substrate distance of 4~cm.
We employed a focused ion beam scanning electron microscopy (FIB-SEM) system (ThermoFisher Scios II) with a Ga$^{+}$ source for nanostructuring of the deposited films in a multistep process using ion beam currents ranging from 300~pA down to 1.5~pA, e.g. for shaping the finest details like the ${\text{Eu}^{3+}:\text{Y}_2\text{O}_3}$ nanopillar. The same FIB-SEM machine was used to acquire the SEM images of the fabricated structures (in Fig.~\ref{fig:5samples}).

\subsection{Computational details}

For the FDTD simulations, we considered an apertured nanopillar setup, close to the manufactured specimen (Fig.~\ref{fig:5samples}a-b), featuring a 220~nm-thick metal film.
\begin{figure}[tb]
    \centering
   \includegraphics[width=1.0\linewidth]{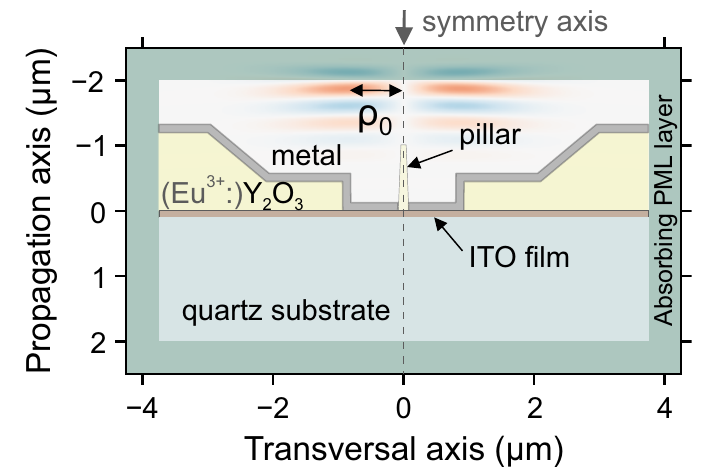}
    \caption{Illustration of the full simulation setup for the apertured nanopillar.
    The central ${\text{Eu}^{3+}:\text{Y}_2\text{O}_3}$ nanopillar (yellow) has the shape of a truncated cone and is surrounded   by a metal layer (dark grey) at the bottom.
    At the lateral parts of the simulation box, we indicated the ${\text{Eu}^{3+}:\text{Y}_2\text{O}_3}$ bulk material (yellow) with a metal layer (dark grey) covering it.
    Underneath the described structure is an ITO layer (brown) deposited on bulk fused quartz (turquoise).
    }
    \label{fig:comp_setup}
\end{figure}
The material description in our simulation setup is based on the Drude--Lorentz oscillator model\cite{1999_jackson_Classical} and tabulated material values from the MEEP materials library\cite{2010_oskooi_Meep} for fused quartz, aluminum, ITO, and yttrium oxide that reproduce empirical values for the complex refractive index\cite{2010_oskooi_Meep}.
Although the experimental pulse duration is in the few-hundred~fs regime, we employed a short, symmetric sin$^2$-laser pulse with a duration of $20~\text{fs}$ in our simulations to reduce their computational cost.
The pulse is polarized purely in the azimuthal direction ($\vec{e}_\phi$) corresponding to an electric field: $\vec{E}(\rho, t)$~=~$E_0\cdot \text{exp}[{\text{i}k \left(z-ct\right)}]\cdot\sqrt{2}\rho/w_0\cdot\text{exp}[-\rho^2/w_0^2]\cdot\vec{e}_\phi$, with the wavenumber $k$ corresponding to the MD transition of ${\text{Eu}^{3+}:\text{Y}_2\text{O}_3}$ with $\lambda$~=~527.5~nm.
We want to emphasize that an increase of the laser pulse duration even beyond some 100~fs will not yield qualitatively different results in our simulations, because thermal effects in the sample materials are assumed to be negligible in the MEEP code.
In accordance to the experiment, we selected an EF peak amplitude of $E_\text{max}=E_0/\sqrt{\text{e}}=0.4$~GV/m (equivalent to $E_0=$0.66~eV and a peak intensity of $2.1\times10^{10}\,\text{W/cm}^2$) and the minimum waist parameter was set to the experimental value of $w_0$~=~1.2~$\mu$m.
The cylindrically symmetric simulation setup considers a volume spanning 7.5~$\mu$m in the transverse and 4.0~$\mu$m in the longitudinal direction along the propagation axis with an omnidirectional spatial resolution of 10~nm.
As can be seen in Fig.~\ref{fig:comp_setup}, the edges of the simulation setup were equipped with a 0.5~$\mu$m artificial absorbing PML layer\cite{2005_taflove_Computational} to prevent spurious back reflections at the box edges.
Within our simulations, the APB pulse was propagated for 60~fs with a temporal resolution of 5~as leading to a Courant factor\cite{1966_yee_Numerical} of less than 0.15 and hence well satisfying the Courant-Friedrichs-Lewy (CFL) condition for numerical stability.

\begin{acknowledgement}
This research was funded in whole or in part by the Austrian Science Fund FWF (grant DOI 10.55776/ZK91 ``iStOMPS''). 
For open access purposes, the authors have applied a CC-BY public copyright license to any author accepted manuscript version arising from this submission. 
The computational results have been achieved using the Austrian Scientific Computing (ASC) infrastructure. 

The authors thank Dariusz Pysz and Ryszard Buczy\'nski for providing the samples of the HC-ARF, and Sarah Pulikottil Alex and Ign\'ac Bug\'ar for their help with the experimental tests of the HC-ARF.  
\end{acknowledgement}

 \begin{suppinfo}
 The following files are available free of charge.
 \begin{itemize}
   \item Spatiotemporal evolution of MFs and EFs for the apertured pillar structure for different polarization directions (longitudinal, azimuthal, radial) (MP4)
 \end{itemize}

 \end{suppinfo}

\providecommand{\latin}[1]{#1}
\makeatletter
\providecommand{\doi}
  {\begingroup\let\do\@makeother\dospecials
  \catcode`\{=1 \catcode`\}=2 \doi@aux}
\providecommand{\doi@aux}[1]{\endgroup\texttt{#1}}
\makeatother
\providecommand*\mcitethebibliography{\thebibliography}
\csname @ifundefined\endcsname{endmcitethebibliography}  {\let\endmcitethebibliography\endthebibliography}{}

\end{document}